# Antiferromagnetic real-space configuration probed by x-ray orbital angular momentum phase dichroism

Margaret R. McCarter,[1] Ahmad I. U. Saleheen,[1] Arnab Singh,[2] Ryan Tumbleson,[1,3] Justin S. Woods,[4,5] Anton S. Tremsin,[6] Andreas Scholl,[1] Lance E. De Long,[4] J. Todd Hastings,[7] Sophie A. Morley,[1] Sujoy Roy[1*]

[1] Advanced Light Source, Lawrence Berkeley National Laboratory, Berkeley, California 94720, USA.
[2] Materials Science Division, Lawrence Berkeley National Laboratory, Berkeley, California 94720, USA.
[3] Department of Physics, University of California, Santa Cruz, California 95064, USA.
[4] Department of Physics and Astronomy, University of Kentucky, Lexington, Kentucky 40506, USA.
[5] Materials Science Division, Argonne National Laboratory, Lemont, Illinois 60439, USA.
[6] Space Sciences Laboratory, University of California, Berkeley, California 94720, USA.
[7] Department of Electrical and Computer Engineering, University of Kentucky, Lexington, Kentucky 40506, USA.
[*] sroy@lbl.gov

**Abstract**

X-ray beams with orbital angular momentum (OAM) are an up-and-coming tool for x-ray characterization techniques. Beams with OAM have an azimuthally varying phase that leads to a gradient of the light field. New material properties can be probed by utilizing the unique phase structure of an OAM beam. Here, we demonstrate a novel type of phase dichroism in resonant diffraction from an artificial antiferromagnet with a topological defect. The scattered OAM beam has circular dichroism whose sign is coupled to the phase of the beam, which reveals the real-space configuration of the antiferromagnetic ground state. Thermal cycling of the artificial antiferromagnet can change the ground state, as indicated by the changing phase dichroism. These results exemplify the potential of OAM beams to probe matter in a way that is inaccessible using typical x-ray techniques.

# MAIN TEXT

## Introduction

X-ray-matter interactions are central to advanced characterization techniques such as x-ray absorption spectroscopy (XAS) and resonant x-ray scattering (RXS), which have transformed our understanding of magnetic, electronic, and orbital ordering phenomena in materials. Typically, these measurements are done by varying the incident x-ray energy and polarization, giving rise to element-specific resonance conditions that enhance signals from magnetic and orbital ordering and dichroic effects. These sensitive techniques can reveal non-trivial spin textures or nano-ordered phases that are difficult to probe with other methods. Some examples include multipolar order parameters (*1–5*), charge ordering in high-temperature superconductors (*6*), and ordering in strongly correlated electron systems (*7,8*) and other quantum materials (*9,10*).

Complex materials such as topological insulators, antiferromagnets, and complex oxides are increasingly explored for technological applications. Antiferromagnets are particularly interesting for spintronic applications. Due to the antiparallel spin arrangement of the two sublattices, antiferromagnets have a net zero magnetization. They have no stray magnetic field, can be used at high frequencies, and are not easily susceptible to external magnetic fields. In ferromagnetic materials, the magnetization can be easily studied, due to the net magnetization that gives rise to x-ray magnetic circular dichroism (XMCD) in XAS. However, the absence of net magnetization in an antiferromagnet usually leads to zero XMCD. Nevertheless, in the case of chiral spin textures (*11–14*), non-zero XMCD can occur. Scattering techniques can probe the staggered moments of an antiferromagnet and are one of the most direct ways of measuring them. In particular, the azimuthal angle or polarization dependence of RXS can measure domains in antiferromagnets with spin spirals (*5,15–19*) or chiral order (*20*). Collinear antiferromagnets can be studied using x-ray magnetic linear dichroism that occurs in XAS and RXS. However, in general, no techniques exist to directly probe the real-space, ground-state configuration of an antiferromagnet or the magnetization of a specific sublattice.

A possible route to advance these techniques is by discovering a new kind of dichroism that takes advantage of the orbital angular momentum (OAM) of x-rays together with the polarization. Here, we show how RXS of x-rays with OAM can be used to distinguish between the two degenerate ground states of an artificial antiferromagnetic lattice. Right- and left-circularly polarized x-rays scatter differently from an antiferromagnet, depending on the phase of the OAM beam, and the resultant dichroism provides information on the magnetic ground state. This new phase dichroism could lead to novel ways to probe materials via interactions with x-ray OAM beams.

OAM is a property of light beams for which the wavefront forms a helix along the propagation direction and the phase varies azimuthally, as shown in Fig. 1A. Due to its changing phase, an OAM beam has an electric field gradient, which can be expected to lead to unique light-matter interactions. Beams with OAM have already found many practical uses in the optical wavelength regime, such as in optical tweezers and subwavelength imaging (*21–27*), and potential exists for future applications in optical communications and quantum optics (*21–24,28*).

However, exploitation of OAM in x-ray beams is a relatively novel field (*29–35*). The generation of x-ray OAM beams has been investigated (*33–36*), but the interaction between x-rays carrying OAM and matter (*37*) has only recently been studied experimentally (*32,38,39*). Potential applications include x-ray holography (*30,31*), ptychography (*36*), and microscopy (*40*).

The electric field gradient of OAM beams could lead to unique dichroic effects in x-ray absorption spectroscopy, such as enhanced sensitivity to electric quadrupole transitions (*37,41*) and molecular chirality (*32,38*). Additionally, because of their helical phase structure, OAM beams have a non-trivial topology, so they could be used to characterize magnetic skyrmions (*30*), vortices (*42,43*), or other magnetic and electric topological textures (*33*). While theoretical predictions have been made (*37,38*), so far experimental verification of the spectroscopic applications of x-ray OAM beams is very limited (*32*).

One method to generate x-ray beams with OAM is to use an artificial antiferromagnet with a built-in topological defect. An artificial antiferromagnet can be created by fabricating a square array of dipole-coupled nanomagnets. The nanomagnets are rectangular in shape so that their strong shape anisotropy constrains the magnetization to align with the long axis of the nanomagnet, thereby creating an analog of an Ising system. Due to the asymmetric interaction between nearest neighbors, the system exhibits an antiferromagnetic ground state. Nanofabrication techniques permit one to tailor the thickness and spacing of the nanomagnets, which enables control of the magnetic transition temperature. Artificial antiferromagnets are widely studied as artificial spin ice systems (*44,45*), for understanding thermal fluctuations in metamaterials (*46–50*), and as candidates for magnonics applications (*50–52*).

RXS from an antiferromagnetic square nano-array gives rise to magnetic diffraction (*34,49,53*). It has been shown that x-rays with OAM are created when diffracted from an artificial antiferromagnetic lattice with a topological edge defect (*34*). Interaction with the defect imparts its topology to the beam, leading to scattered beams with OAM related to the topological charge of the defect. Furthermore, antiferromagnetic ordering coupled to OAM leads to interesting interference effects that can be observed using circularly polarized x-rays (*34*). This study shows how these effects can be utilized to determine the real-space, ground-state configuration of an artificial antiferromagnet.

**Results**

*Photoemission electron microscopy*

An edge defect in a lattice can be characterized by its topological charge $\mathbb{Z}$, which is equal to the number of edge dislocations. We fabricated $\mathbb{Z}2$ artificial antiferromagnetic lattices (with topological charge of 2) using electron-beam lithography and lift-off on ferromagnetic permalloy ($Fe_{0.8}Ni_{0.2}$). Real-space imaging was done using scanning electron microscopy (SEM), as seen in Fig. 1B. The lattice is a 10 x 10 μm² square array made of 470 nm x 170 nm nano-islands with a lattice spacing $a = 700$ nm; the $\mathbb{Z}2$ topological defect is located at the array center. The defect can be quantified by the Burgers vector $\vec{t} = 2a\hat{x}$, with $\hat{x}$ defined as the direction parallel to the edge dislocations.

X-ray magnetic circular dichroism photoemission electron microscopy (XMCD PEEM) was used to investigate the magnetic configuration of the antiferromagnetic state. The beam energy was tuned to the Fe $L_3$ edge (707.6 eV) to obtain magnetic contrast. Images were taken for right- and left-circularly polarized x-rays incident on the sample along the [-1,1] direction (at a 135° angle with respect to the Burgers vector). An XMCD image is created by taking the difference between the images for right- and left-circularly polarized x-rays, which results in bright and dark islands, as seen in Fig. 1B. Bright and dark islands have a component of in-plane magnetization parallel or antiparallel, respectively, to the in-plane direction of the incident x-ray beam. This information can be used to extract the magnetic configuration (see Supplementary Materials for more information).

Since the magnetic lattice spacing is twice that of the structural lattice, the magnetic configuration effectively has one edge dislocation with a topological charge $\mathbb{Z}_m = \pm 1$, which depends on the direction of magnetization surrounding the defect. Thus, our sample array contains a structural and a magnetic defect with different topological charge. We define the magnetic configuration in Fig. 1B as the $\mathbb{Z}_m = +1$ defect, with a net clockwise magnetization surrounding the defect. A defect with a net counterclockwise winding of the magnetization has $\mathbb{Z}_m = -1$. Since the $\mathbb{Z}_m = \pm 1$ states are degenerate, it should be possible to stabilize both ground states with an equal probability if the artificial antiferromagnet is cycled multiple times above and below the magnetic transition temperature.

*Coherent resonant x-ray scattering*

We have characterized the two antiferromagnetic ground states using coherent RXS tuned at the Fe $L_3$ edge (707.6 eV) with the scattering geometry shown in Fig. 1C. Circularly polarized x-rays are incident on the sample at an angle $\theta = 9°$ to the array plane. The Burgers vector of the edge defect is parallel to the in-plane direction of the incident x-ray beam (defined here as the $\hat{x}$-direction). The lattice gives rise to a diffraction pattern with different peaks arising from the structural and magnetic lattices, respectively. When scattering from the artificial antiferromagnet, structural peaks appear at $(H, K)$ when $H$ and $K$ are even integers, whereas magnetic peaks appear at $(H, K)$ when $H$ and $K$ are odd integers. This is because the periodicity of the antiferromagnetic lattice is twice that of the structural lattice. Any peak with a component of its scattering vector in the same direction as the Burgers vector (*i.e.*, any peak with $H \neq 0$) will have a non-zero OAM (*34*), whose OAM value is equal to $\ell \hbar$, where $\ell$ is equal to the diffraction peak order $H$.

The diffraction pattern in Fig. 2A is the sum of scattering for right- and left-circularly polarized incident x-rays. The pattern has a set of magnetic peaks at $H = \pm 1$ and structural peaks at $H = 0$. The magnetic peaks have OAM, leading to characteristic peaks with a central dark spot due to the phase singularity. The structural peaks at $H = 0$ have no OAM, because the scattering is perpendicular to the Burgers vector (*i.e.*, the peaks are scattered in the $\hat{y}$-direction, whereas the Burgers vector is along the $\hat{x}$-direction). Additionally, the scattering vector is related to the sign of the OAM. The peaks at $H = \pm 1$ have OAM values equal to $\ell = \pm 1$, meaning that the phase wraps clockwise as seen in Fig. 1A for $H = +1$, or counterclockwise for $H = -1$.

It has been shown (*34*) that the magnetic diffraction peaks from the $\mathbb{Z}2$ lattice differ for right- versus left-circularly polarized incident x-rays, giving rise to a circular dichroism at the peak. This can be quantified by the circular asymmetry, which is defined as:

$$asymmetry = \frac{I_{rc} - I_{lc}}{I_{rc} + I_{lc}}$$

where $I_{rc}$ and $I_{lc}$ are the scattered intensities for right- and left-circularly polarized x-rays. The sign of the asymmetry has a distinct pattern, as shown in Fig. 2B, which is related to the phase of the OAM beam. The asymmetry is half positive and half negative at each peak, so it switches sign when the OAM beam phase changes by π. The asymmetry pattern also reverses sign upon changing from $H = +1$ to $H = -1$. The dichroism is then linked to the OAM of the beam, since changing from $H = +1$ to $H = -1$ also changes the OAM value of the beam from $\ell = +1$ to $\ell = -1$.

For the $H = +1$ peaks, the asymmetry implies that the right- (left-) circularly polarized OAM beam scatters with higher intensity to the top (bottom) half of the ring-shaped diffracted beam. In other words, the spin-orbit angular momentum of the light and its interaction with the material's spin degree of freedom introduces a term in the scattering cross-section that depends on the phase progression of the scattered beam. We note that

right- and left-circularly polarized Gaussian beams should scatter with the same intensity at the antiferromagnetic Bragg peaks for an antiferromagnet with no defect (*i.e.*, for scattered beams with $\ell = 0$). This is the reason that an ordered antiferromagnet does not usually give rise to circular dichroism.

Since the $H = \pm 1$ peaks arise due to scattering from the antiferromagnetic ordering, the diffraction peaks and dichroism should disappear above the transition temperature, which is about $T_N \approx 380$ K for this artificial antiferromagnet (*34*). To confirm this, we performed temperature-dependent measurements, as shown in Fig. 2C. At room temperature, a line profile through an $H = +1$ diffraction peak clearly shows the half-positive, half-negative dichroism distribution. The dichroism persists to at least 320 K. At 380 K, there is no longer any dichroism, confirming that the peaks and dichroism disappear once the antiferromagnetic state is suppressed by thermal fluctuations.

*Magnetic scattering calculations*

To investigate how RXS can be used to distinguish between the two degenerate ground states of the antiferromagnet, we use resonant scattering calculations for the different states. In RXS, the resonant scattering amplitude from the $n^{\text{th}}$ scatterer in a magnetic system is usually expressed to first order in the magnetization using the electric-dipole approximation:

$$f_n = f_0(E)\vec{\varepsilon}'^* \cdot \vec{\varepsilon} - if_1(E)(\vec{\varepsilon}'^* \times \vec{\varepsilon}) \cdot \vec{m}_n$$

where $f_0$ and $f_1$ are energy-dependent constants, $\vec{\varepsilon}'$ and $\vec{\varepsilon}$ are the polarization of the scattered and incident x-rays, respectively, and $\vec{m}_n$ is the magnetization direction. The intensity $I$ of scattering is equal to the sum over all scatterers:

$$I = \sum_n \left| f_n e^{i\vec{q}\cdot\vec{r}_n} \right|^2$$

where $\vec{q}$ is the scattering vector ($\vec{q} = \vec{k}' - \vec{k}$) and $\vec{r}_n$ is the position of the $n^{\text{th}}$ scatterer. We calculated the scattered intensity for incident right- and left-circularly polarized x-rays and determined the circular dichroism (details are available in the Supplementary Materials).

Scattering from the 10 x 10 μm² nanomagnet array was calculated, and the resulting intensity profile is shown in Fig. 3A for the sum of right- and left-circularly polarized incident x-rays. The experimental scattering pattern is reproduced well, with OAM beams with a central dark spot due to the phase singularity appearing at the expected $(H, K)$ values. Differences between the experimental and theoretical beam shape are due to additional interference effects that arise from the coherent x-ray beam used in the experiment (see Supplementary Materials for more information).

Next, we calculated the difference in scattered intensity for right- or left-circularly polarized, incident x-rays. We simulated the two possible magnetic ground states of the artificial antiferromagnet, as shown in Fig. 3B, with magnetic defects of topological charge $\mathbb{Z}_m = \pm 1$. Figure 3C shows the resulting circular dichroism patterns. The $\mathbb{Z}_m = +1$ configuration matches the experimentally observed pattern in Fig. 2B. The circular dichroism pattern reverses for the other ground state, showing that this procedure can be used to distinguish between the two antiferromagnetic ground states.

The phase of the scattered OAM beams at $H = \pm 1$ are shown in Fig. 3D. The opposite peaks have a phase that wraps either clockwise or counterclockwise with respect to the beam propagation direction. Considering the $H = +1$ peak, which wraps clockwise, we can see the relationship between the dichroism and the winding magnetization around the defect. When the defect winds in the same (opposite) direction as the OAM phase, the right-circularly polarized beam is scattered preferentially up (down), and similarly the left-circularly polarized beam is scattered preferentially down (up), giving rise to the dichroism patterns in Fig. 3C.

The two antiferromagnetic ground states are degenerate, so they should form with equal probability if the antiferromagnet is heated above its transition temperature and then returned to room temperature. After cooling, a flip of the dichroism pattern would indicate a change in the antiferromagnetic ground state, according to the theoretical calculations. In order to test this experimentally, we performed sequential measurements on an artificial antiferromagnet as it was heated to 380 K and cooled back to room temperature. As shown in Fig. 4, the room-temperature dichroism pattern forms in both configurations with about 50-50 probability, which is expected for random thermal switching between two degenerate ground states.

**Discussion**

The phase dichroism that we report here can be observed if the scattering pattern is spatially resolved at the detector. The integrated intensity of an antiferromagnetic Bragg peak is the same for both the right- and left-circularly polarized light. Thus, our result is consistent with polarization-dependent resonant scattering. As anticipated, it is difficult to study antiferromagnetic domains due to their zero net magnetization, particularly in the case of phase domains where the order parameter remains the same, but the phase is shifted by 180 degrees (*19*). For the case of scattering from the artificial antiferromagnet presented here, circular dichroism related to the phase of the scattered OAM beam can be used to determine the real-space antiferromagnetic configuration. It is likely that with a beam size smaller than antiferromagnetic domains, it will be possible to determine the precise ground state of antiferromagnetic domains by measuring their phase dichroism.

In our experiment, complete understanding of an antiferromagnetic ground state configuration implies we can access sufficient information to characterize the specific antiferromagnetic sublattices (*i.e.*, which domains have magnetization parallel, antiparallel, or perpendicular to the x-ray beam). Our results indicate x-ray beams with OAM can be used to probe a specific sublattice within an antiferromagnet, which could lead to new techniques for imaging antiferromagnetic domains. A promising future avenue will be to first create an OAM beam and then use it to measure resonant diffraction from a traditional antiferromagnet. If the magnetic interaction of the OAM beam is phase-dependent, then circular dichroism in the scattered beam should provide information on the antiferromagnetic configuration. Given the small size of the incident OAM beam, it is also naturally suited for nano-diffraction applications, where it could identify domain walls, chiral defects, or topological defects. Furthermore, if an OAM beam can be used to measure specific spin sublattices, it could also be used as a direct method for measuring spin currents.

Finally, an analogy can be drawn to the photonic spin Hall effect (PSHE). The PSHE describes photons with different circular polarizations (*i.e.*, photons with opposite spin angular momenta) displaced in opposite directions after interacting with a medium with inhomogeneous refractive index. For the PSHE, the photons and spatially varying refractive index play a role similar to charge carriers and the electric potential gradient in the ordinary Hall effect. Experiments demonstrating the PSHE often use metamaterials (structures that are engineered to have properties different than those of the constituent material) to study this spin-orbit interaction of light (*54–57*). In the present case, our artificial antiferromagnet also acts as a metamaterial, and the interplay between spin and orbital angular momentum leads to a process where different circular polarizations are scattered asymmetrically.

## Materials and Methods

### Sample fabrication
The artificial antiferromagnetic square lattices were made of a 2.4-nm-thick layer of permalloy ($Ni_{0.8}Fe_{0.2}$) capped with 1.5 nm of aluminum. The permalloy thickness was chosen to create an artificial lattice that is antiferromagnetic at room temperature. The transition temperature for this thickness is ~380 K. To pattern the nanomagnets, electron beam lithography was used on a Si substrate with a bilayer of PMMA positive resist. Then, electron beam evaporation was used to deposit the permalloy and aluminum layers. Deposition was done at a rate of 0.2 Å/s and a pressure of $10^{-7}$ Torr.

### Photoemission electron microscopy
PEEM characterization was done at the PEEM-3 beamline 11.0.1.1 at the Advanced Light Source. Measurements were taken at the Fe $L_3$ edge with right- and left-circularly polarized x-rays incident along the [-11] in-plane direction of the sample at 30° grazing incidence.

### Coherent resonant x-ray scattering
Experiments were performed at the COSMIC scattering beamline 7.0.1.1 at the Advanced Light Source. To create a coherent beam, a 7-μm circular pinhole was used. Measurements were taken at the Fe $L_3$ edge with right- and left-circularly polarized x-rays. The scattering patterns were measured with a Timepix-based area detector (*58–60*). The incident angle of x-rays on the sample was $\theta = 9°$, and the detector was located at $2\theta = 18°$. A beam stop was used to block the bright specular reflection so that the diffraction pattern from the artificial antiferromagnet could be measured. When calculating the circular asymmetry, background regions with low intensity can give rise to large asymmetry values due to the small denominator in the calculation. Because of this, an appropriate intensity threshold value was chosen to set the asymmetry to zero in low intensity regions so that the real asymmetry occurring at the diffraction peaks is clear.

**Acknowledgments**

**Funding:** This work was supported by the Laboratory Directed Research and Development Program of Lawrence Berkeley National Laboratory under U.S. Department of Energy Contract No. DE-AC02-05CH11231. This work used Timepix based soft x-ray detector, development of which is supported by DOE through award RoyTimepixDetector. This research used resources of the Advanced Light Source, a U.S. DOE Office of Science User Facility under contract no. DE-AC02-05CH11231. Work performed at the Center for Nanoscale Materials, a U.S. Department of Energy Office of Science User Facility, was supported by the U.S. DOE, Office of Basic Energy Sciences, under Contract No. DE-AC02-06CH11357. Sample characterization was performed in the Material Science Division at Argonne National laboratory and supported by the U.S. Department of Energy, Office of Science, Basic Energy Sciences, Materials Sciences and Engineering Division. The work is supported by the U.S. Department of Energy, Office of Science, Office of Basic Energy Sciences under Award No. DE-SC0016519. This work was performed in part at the University of Kentucky Center for Nanoscale Science and Engineering, Electron Microscopy Center, and Center for Advanced Materials, members of the National Nanotechnology Coordinated Infrastructure (NNCI), which is supported by the National Science Foundation (NNCI-2025075).

**Author contributions:**

Conceptualization: SR, MRM
Sample fabrication: JSW

Scanning electron microscopy measurements: JSW, SAM
Photoemission electron microscopy measurements: MRM, SR, AScholl
Coherent resonant x-ray scattering measurements: MRM, AIUS, ASingh, RT, SAM
Resonant scattering calculations: MRM
Writing—original draft: MRM, SR
Writing—review & editing: MRM, AIUS, ASingh, RT, JSW, AST, AScholl, LEDL, JTH, SAM, SR

**Competing interests:** Authors declare that they have no competing interests.

**Data and materials availability:** All data needed to evaluate the conclusions in the paper are present in the paper and/or the Supplementary Materials. Additional data related to this paper may be requested from the authors.

**Figures and Tables**

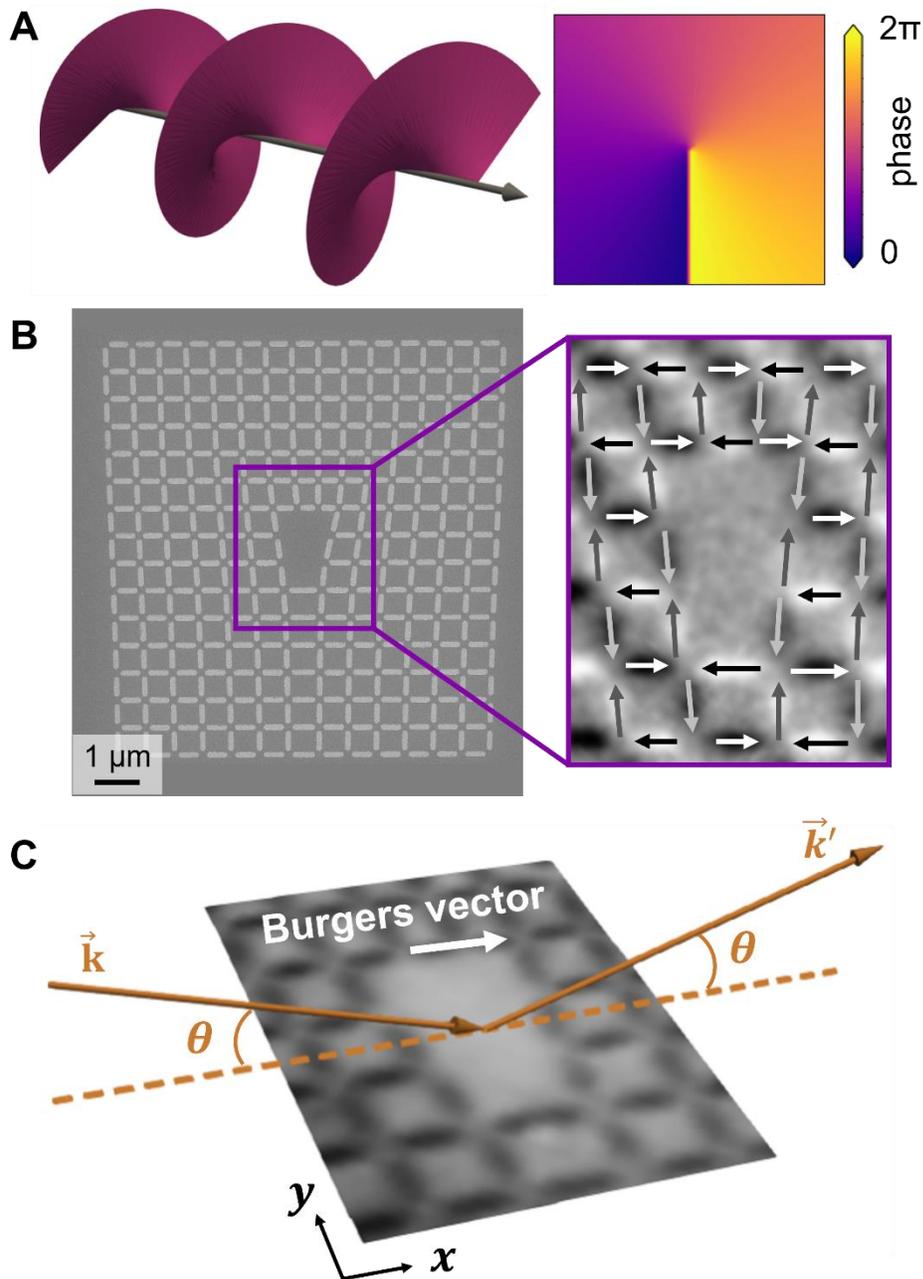

**Fig. 1. Experimental details of orbital angular momentum beams, photoemission electron microscopy, and resonant scattering.** (**A**) Left: The helical phase front of a beam carrying OAM, where the helix represents a surface of constant phase. Right: The cross-section through an OAM beam has a phase that varies azimuthally. Pictured is a beam with topological charge $\ell = +1$, which has a phase that wraps once clockwise around the propagation direction. (**B**) Scanning electron microscopy image of an artificial antiferromagnet with a topological defect (left), and XMCD PEEM (right) showing the antiferromagnetic configuration around the defect. (**C**) The geometry for resonant scattering from the artificial antiferromagnet. The in-plane direction of the incident beam is parallel to the defect Burgers vector.

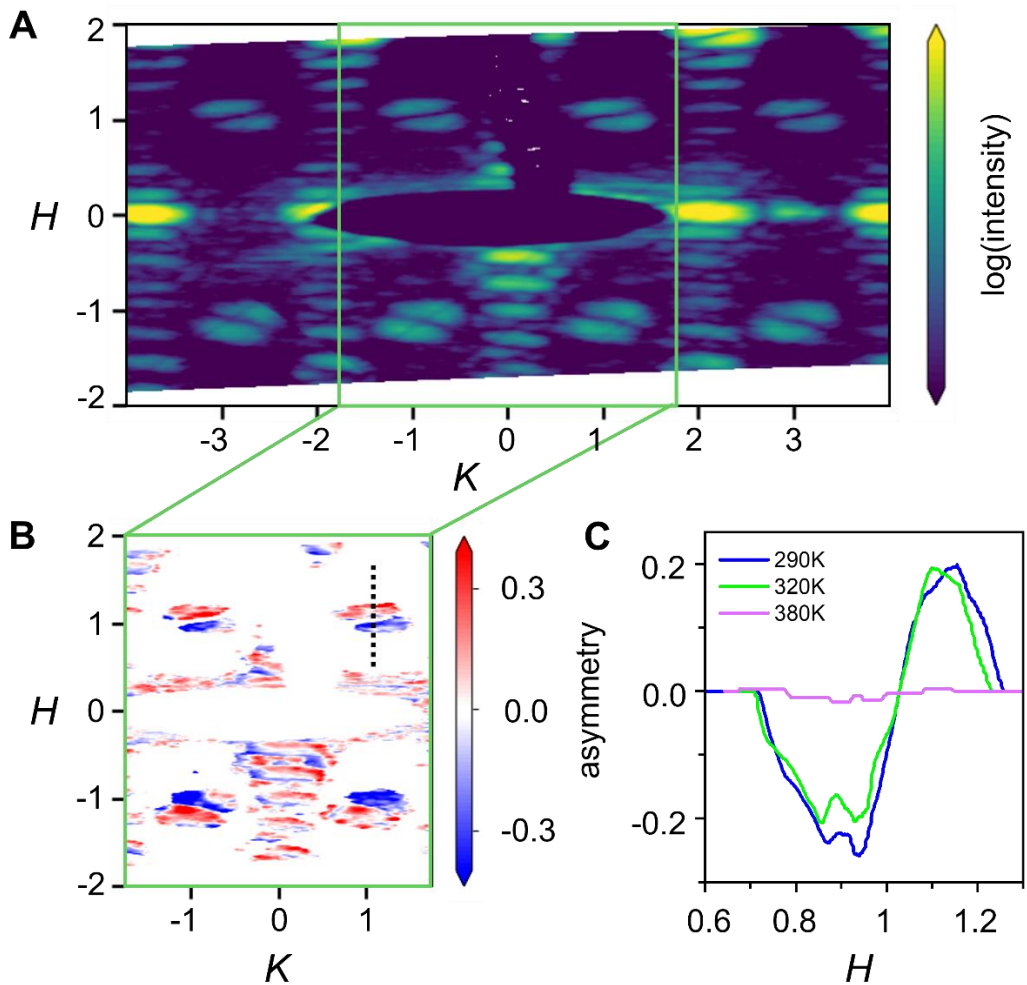

**Fig. 2. Resonant x-ray scattering from an artificial antiferromagnet.** (**A**) The sum of diffraction patterns for right- and left-circularly polarized incident x-rays plotted on a logarithmic scale. The pattern has magnetic diffraction peaks at $H = \pm 1$ from the antiferromagnetic ordering. (**B**) The circular asymmetry shows circular dichroism at the peaks. Red/blue corresponds to positive/negative circular asymmetry. (**C**) The temperature-dependent circular asymmetry measured along the line marked in **B**. The circular asymmetry disappears upon heating above the antiferromagnetic transition temperature $T_N \approx 380$ K.

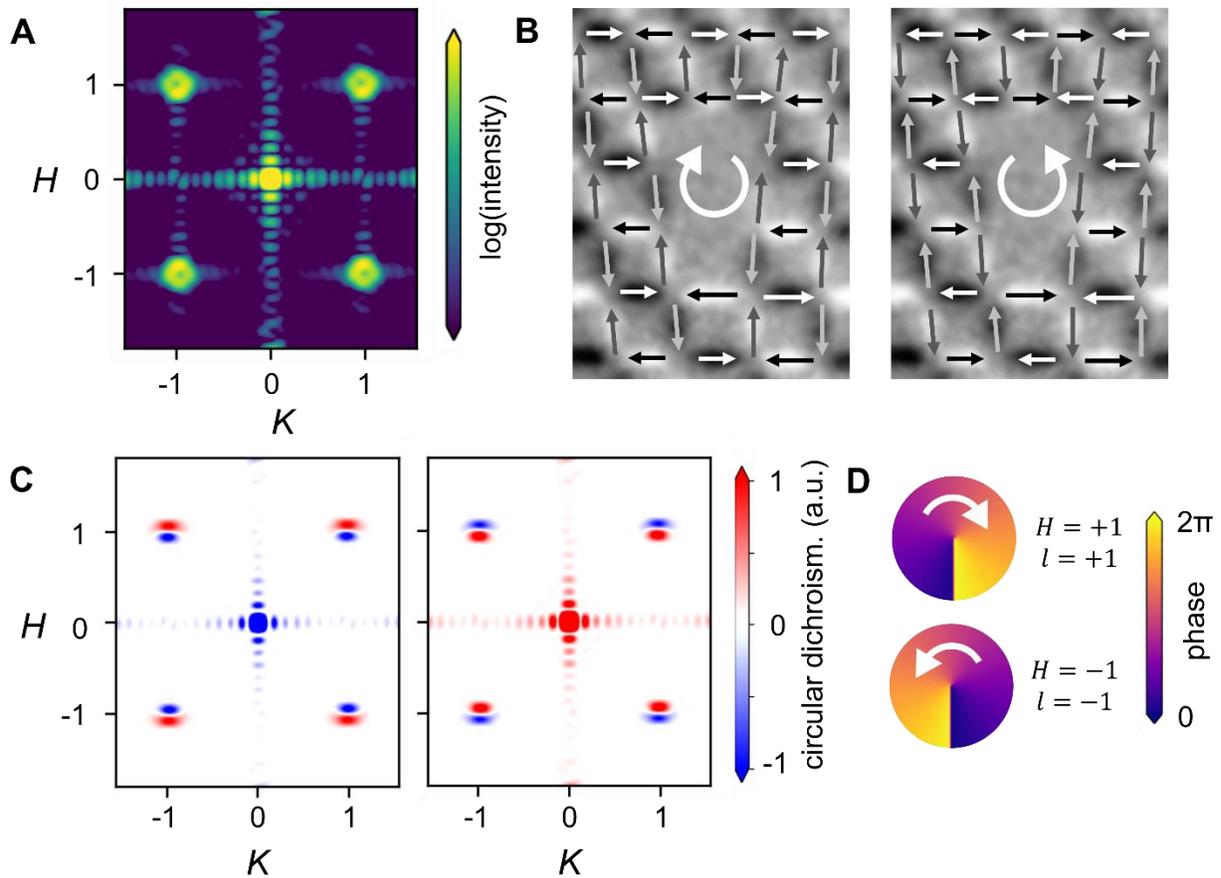

**Fig. 3. Resonant x-ray scattering calculations.** (**A**) The calculated diffraction from the artificial antiferromagnet with a $\mathbb{Z}2$ edge defect. (**B**) The two possible antiferromagnetic ground states with $\mathbb{Z}_m = \pm 1$ overlaid on an XMCD PEEM image. (**C**) The simulated circular dichroism patterns corresponding to the configurations in **B**. (**D**) The phase of the beams for $H = \pm 1$. The direction of the phase gradient (white arrows) interacts differently with the two magnetic defect configurations, giving rise to the two distinct asymmetry patterns in resonant scattering.

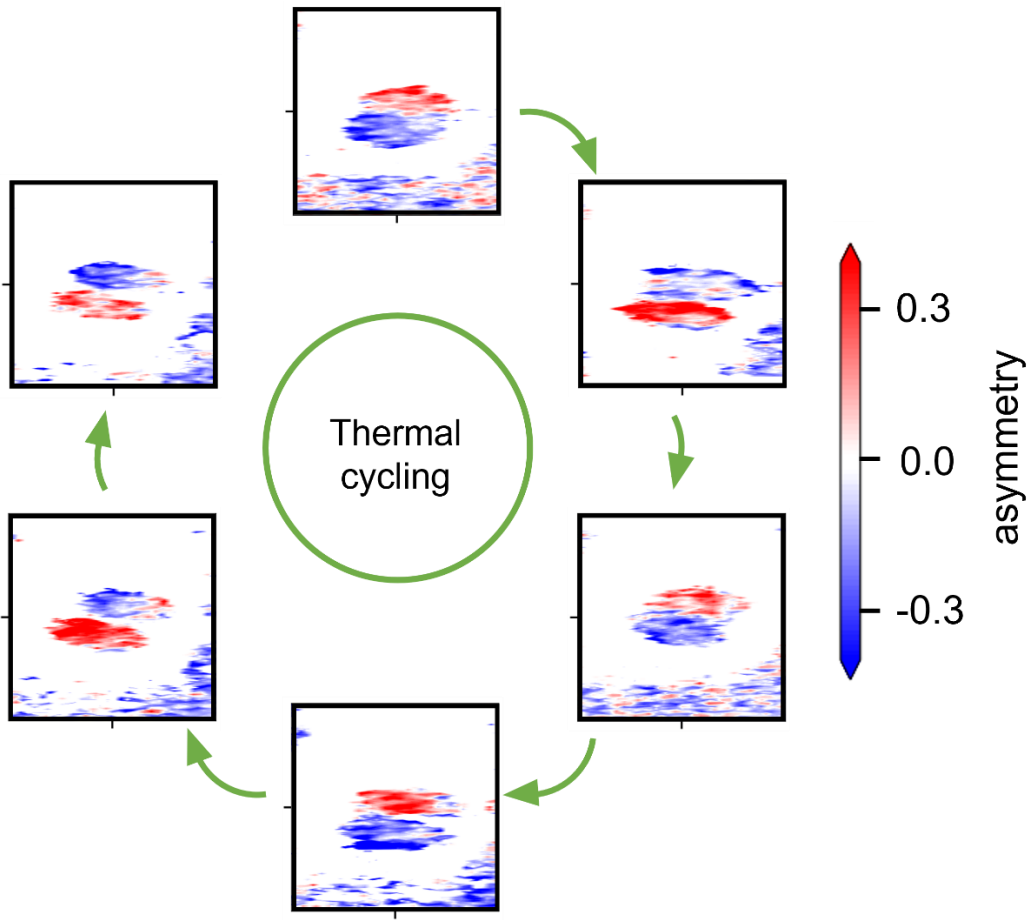

**Fig. 4. Changing the antiferromagnetic ground state with thermal cycling.** The $(H, K) = (1, -1)$ peak is shown. When heated to 380K and cooled back to room temperature, the artificial antiferromagnet randomly forms in one of two antiferromagnetic ground states, as evidenced by the circular asymmetry at the diffraction peak. The room temperature dichroism pattern is shown here after each of six subsequent thermal cycles.

## Supplementary Text

Photoemission electron microscopy

Photoemission electron microscopy (PEEM) with x-ray magnetic circular dichroism (XMCD) can be used to image the artificial antiferromagnetic lattice. Measurements were done with x-rays incident along the [-1,1] direction. Images were taken for right and left circularly polarized incident x-rays. The sum of images for both polarizations is shown in Fig. S1A. This shows the lattice configuration without magnetic contrast. The XMCD images, as shown in Fig. S1B are obtained by taking the difference of the images for right and left circular polarizations. The result shows alternating stripes of bright and dark islands arranged along the [-1,1] direction. The bright (dark) contrast in PEEM XMCD indicates that the islands have a magnetization component parallel (perpendicular) to the direction of the incoming x-rays. Additionally, because of the shape anisotropy of the nanomagnets, the magnetization of each island is constrained to lie along the long axis of the nanomagnet (parallel or antiparallel to the *x*- or *y*-direction). Using this information, the magnetization configuration can be deduced, as illustrated in Fig. S1C and shown in Fig. 1B.

Resonant scattering calculations

The intensity of resonant scattering can be written as:

$$I = \sum_n \left| f_n e^{i\vec{q}\cdot\vec{r}_n} \right|^2$$

where $\vec{q}$ is the scattering vector and $\vec{r}_n$ is the position of the $n^{th}$ scatterer. The scattering vector is defined in terms of the wave vectors of the scattered and incident x-rays as $\vec{q} = \vec{k}' - \vec{k}$, as shown in Fig. S2. In resonant scattering, the resonant scattering amplitude from the $n^{th}$ scatterer in a magnetic system is usually approximated as (*61*):

$$f_{n,\vec{\varepsilon}',\vec{\varepsilon}} = f_0(E)\vec{\varepsilon}'^* \cdot \vec{\varepsilon} - if_1(E)(\vec{\varepsilon}'^* \times \vec{\varepsilon}) \cdot \vec{m}_n$$

where $f_0$ and $f_1$ are energy-dependent constants, $\vec{\varepsilon}'$ and $\vec{\varepsilon}$ are the polarization of the scattered and incident x-rays, respectively, and $\vec{m}_n$ is the magnetic dipole moment. This formulation can be written in terms of the resonant scattering of linearly polarized light (*61*):

$$f_{n,\sigma',\sigma} = f_0(E)$$

$$f_{n,\sigma',\pi} = -if_1(E)m_{n,x}\cos\theta$$

$$f_{n,\pi',\sigma} = if_1(E)m_{n,x}\cos\theta$$

$$f_{n,\pi',\pi} = f_0(E)\cos2\theta + m_{n,y}\sin2\theta$$

where $\sigma$ and $\pi$ represent linear polarization perpendicular or parallel to the scattering plane, respectively. For circularly polarized light, the intensity of scattered light can be rewritten as (*62*):

$$I = \frac{1}{2}\left( |G_{\sigma',\sigma}|^2 + |G_{\pi',\sigma}|^2 \right) + \frac{1}{2}\left( |G_{\pi',\pi}|^2 + |G_{\sigma',\pi}|^2 \right) + P_3 Im\left( G_{\sigma',\pi}^* G_{\sigma',\sigma} + G_{\pi',\pi}^* G_{\pi',\sigma} \right)$$

where:

$$G_{\vec{\varepsilon}',\vec{\varepsilon}} = \sum_n f_{n,\vec{\varepsilon}',\vec{\varepsilon}} e^{i\vec{q}\cdot\vec{r}_n}$$

and $P_3$ is the Stoke's parameter equal to $+1$ and $-1$ for right- or left-circularly polarized x-rays, respectively.

Because the main contributors to magnetic scattering are the islands magnetized along the *x*-direction (*i.e.*, the islands with a component of magnetization along the incident x-ray propagation direction), we performed the summation for the sublattices of the artificial spin ice with $m_{n,x} \neq 0$, represented by the red and blue sublattices in Fig. S2, where red and blue represent magnetization along the $+x$- and $-x$-direction, respectively. Calculations were performed for this magnetization configuration as well as the opposite configuration (with red and blue representing magnetization along the $-x$- and $+x$-direction, respectively). These configurations are also shown in Fig. 3B of the text.

Coherent diffraction effects on the scattered beam shape

Typically, a beam with OAM has a donut-shaped intensity profile. This has been observed in previous x-ray experiments (32). Performing RXS experiments without a pinhole, we also observe donut-shaped beams (Fig. S3A), and we observe the expected circular asymmetry at the peaks (Fig. S3B). The data in the main text (reproduced here in Fig. S3C and S3D) was taken with a 7-µm circular pinhole placed before the sample. This cuts down background intensity from the specular reflection and makes it easier to observe the diffraction peaks and circular asymmetry. However, the introduction of the pinhole also distorts the donut-shape of the beam, due to additional interference effects from the coherence of the beam and/or due to finite-size effects from the reduced beam size. While the intensity profile is affected, the circular asymmetry pattern is not. The opposite asymmetry in Fig. S3B and S3D is due to thermal cycling that took place between the measurements, meaning the artificial antiferromagnetic ground state had changed.

**Fig. S1.**

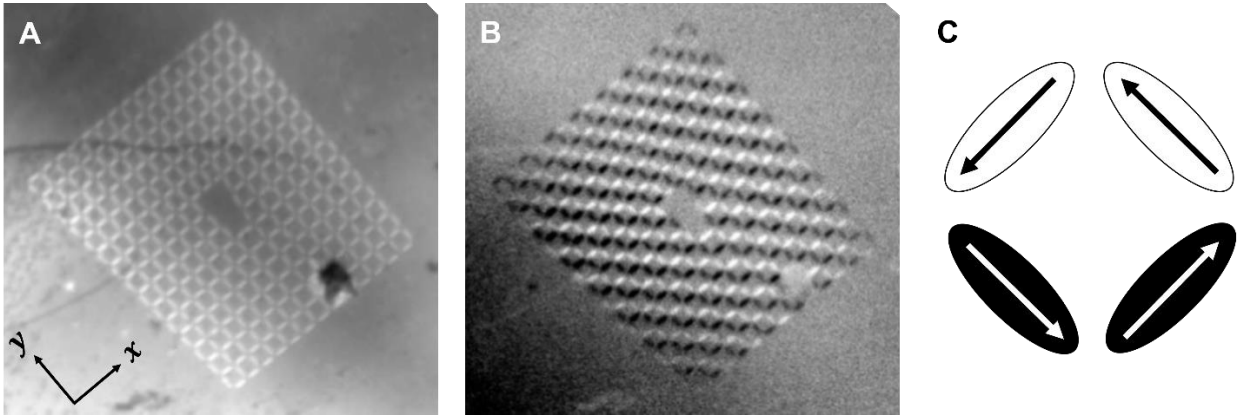

**Fig. S1. PEEM XMCD imaging**. (**A**) The sum of images for right- and left-circularly polarized x-rays. (**B**) The difference image for right- and left-circularly polarized x-rays gives magnetic contrast, with alternating stripes of bright and dark islands representing regions with magnetization parallel or antiparallel to the incident x-ray beam. (**C**) Using the XMCD contrast and knowledge of the shape anisotropy of the nanomagnets, the magnetization direction of each island can be determined.

**Fig. S2.**

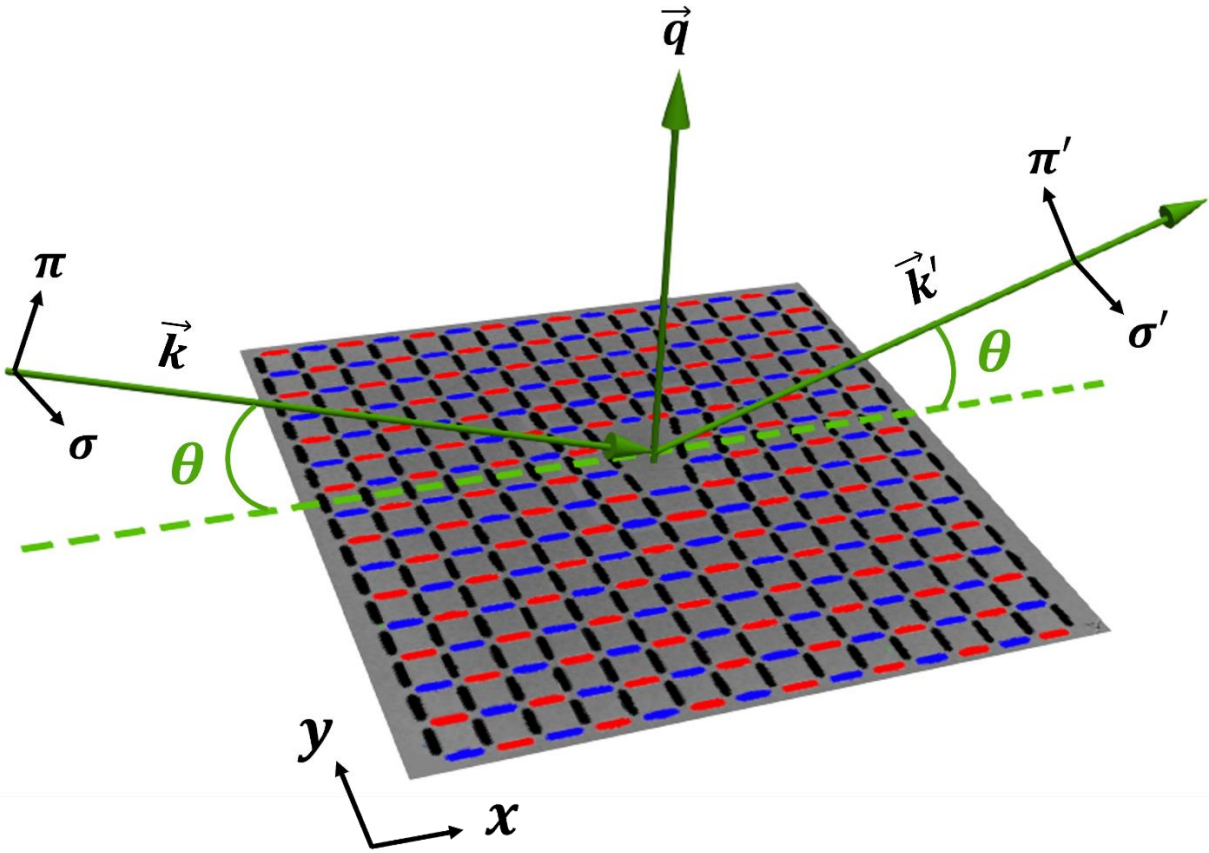

**Fig. S2. Resonant scattering geometry.** The resonant scattering geometry is shown, including the scattering vector $\vec{q}$, the incident and scattered wave vectors $\vec{k}$ and $\vec{k}'$, the incident angle $\theta$, and the polarization directions $\sigma/\pi$ and $\sigma'/\pi'$ for the incident and scattered x-rays, respectively. The red and blue sublattices on the artificial spin ice represent islands with magnetization parallel and antiparallel to the *x*-direction, which were used in the scattering calculations.

**Fig. S3.**

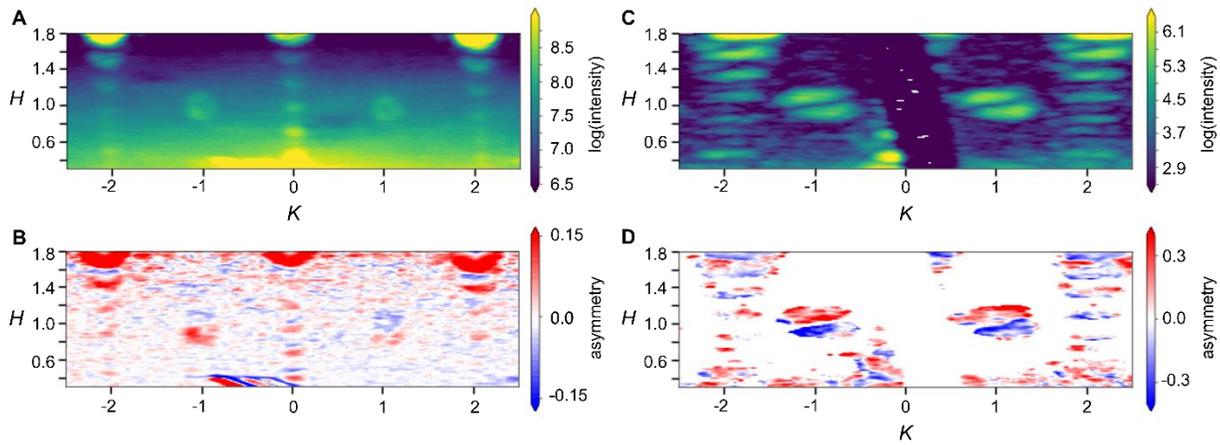

**Fig. S3. Resonant scattering measurements with and without a pinhole**. (**A**) Resonant scattering intensity with no pinhole, which leads to donut-shaped beams. (**B**) The circular asymmetry pattern with no pinhole. (**C**) Resonant scattering intensity with a 7-µm pinhole inserted before the sample, leading to a distorted donut-shaped beam. (**D**) The circular asymmetry with the 7-µm pinhole.